\documentstyle[epsfig]{mn}
\begin{document}
%\input /home/sbgs/latex/BoxedEPS.tex
%\input /home/sbgs/latex/macro.tex
%\SetEPSFDirectory{/scratch/sbgs/figures/hst/}
%\SetRokickiEPSFSpecial
%\HideDisplacementBoxes

%%%%%%%%%%%%%%% MODIFICATION HISTORY %%%%%%%%%%%%%%%%%%%%
%%%%%
%%%%%  Added addresses and slightly altered author lists
%%%%%   Stephen Serjeant Thu Nov 21 --> version 0.2
%%%%%  Changed Nicholas Eaton to N.Eaton. SS Nov 21
%%%%%  
%%%%%  
%%%%%%%%%%%%%%%%%%%%%%%%%%%%%%%%%%%%%%%%%%%%%%%%%%%%%%%%%

\title[Observations of HDF with ISO I]{Observations of the
Hubble Deep Field with the Infrared Space Observatory V: Spectral
Energy Distributions, Starburst Models and Star Formation History}
%\thanks{Based on observations with ISO, an ESA project, with instruments funded by ESA
%Member States (especially the PI countries: France, Germany, the Netherlands and the United Kingdom) and 
%with participation of ISAS and NASA}}
\author[Rowan-Robinson, Mann, Oliver, {\it et al.}]
{M. Rowan-Robinson$^1$, R.G. Mann$^1$, S.J. Oliver$^1$, 
A. Efstathiou$^1$,
N. Eaton$^1$,
\vspace*{0.2cm}\\{\LARGE  
P. Goldschmidt$^1$, 
B. Mobasher$^1$,
S. Serjeant$^1$, T.J. Sumner$^1$,
L. Danese$^2$, 
D. Elbaz$^3$, 
}\vspace*{0.2cm}\\{\LARGE 
A. Franceschini$^4$,
E. Egami$^5$, 
M. Kontizas$^6$, 
A. Lawrence$^7$, 
R. McMahon$^8$, 
}\vspace*{0.2cm}\\{\LARGE 
H.U. Norgaard-Nielsen$^9$, 
I. Perez-Fournon$^{10}$, 
J.I. Gonzalez-Serrano$^{11}$
}\\
$^1$Astrophysics Group, Imperial College London, Blackett Laboratory,
Prince Consort Road, London SW7 2BZ;\\ 
$^2$SISSA, Via Beirut 2-4, Trieste, Italy;\\
$^3$Service d'Astrophysique, Saclay, 91191, Gif-sur-Yvette, Cedex,
France;\\ 
$^4$Osservatorio Astronomico de Padova, Vicolo dell'Osservatorio 5,
I-35 122, Padova, Italy;\\
$^5$Max-Planck-Institut f\"ur Extraterrestrische Physik,
Giessenbachstrasse, D-8046, Garching bei Munchen, Germany;\\
$^6$Astronomical Institute, National Observatory of Athens, P.O.Box
200048, GR-118 10, Athens, Greece;\\
$^7$Institute for Astronomy, University of Edinburgh, Blackford Hill,
Edinburgh, EH9 3HJ;\\
$^8$Institute of Astronomy, The Observatories, Madingley Road,
Cambridge, CB3 0HA;\\
$^9$Danish Space Research Institute, Gl. Lundtoftevej 7, DK-2800
Lyngby, Copenhagen, Denmark;\\
$^{10}$Instituto Astronomico de Canarias, Via Lactea, E-38200 La
Laguna, Tenerife, Canary Islands, Spain;\\
$^{11}$Instituto de Fisica de Cantabria, Santander, Spain\\
}
\maketitle
\begin{abstract}
We have modelled the spectral energy distributions of the 13 HDF galaxies reliably detected by ISO. 
For 2 galaxies the emission detected by ISO is
consistent with being starlight or the infrared 'cirrus' in the galaxies.  For the remaining 11 galaxies there
is a clear mid-infrared excess, which we interpret as emission from dust associated with a strong
starburst.  10 of these galaxies are spirals or interacting pairs, while the remaining one is an elliptical
with a prominent nucleus and broad emission lines. 

We give a new discussion of how the star formation rate can be deduced from the far infrared luminosity
and derive star formation rates for these galaxies of 8-1000 $\phi M_{\sun}$ per yr, where $\phi$ takes account of the
uncertainty in the initial mass function.  The HDF galaxies detected by ISO
are clearly forming stars at a prodigious rate compared with nearby normal galaxies. We discuss the
implications of our detections for the
history of star and heavy element formation in the universe.  Although uncertainties
in the calibration, reliability of source detection, associations, and starburst models remain, 
it is clear that dust plays an important role in star formation out to
redshift 1 at least.

\end{abstract}

%\large

\section{Introduction} 

Because of its great depth, high resolution, and the intensive follow-up which has been
carried out in it, the Hubble Deep Field (HDF) is an exceptional resource for cosmological studies.
The central area of the HDF consists of 5 square arcmin of sky.  It was imaged by the
Hubble Space Telescope on 150 orbits in December 1995 and reaches to at least 29th
magnitude in I (800nm), V (600nm) and B (450nm), and to 27th magnitude in U (300nm)
(Williams et al 1996).
We were successful in bidding for Director's Time on the Infrared Space Observatory
(ISO) and were awarded a total of 12.5 hours to map the HDF with ISO-CAM in the LW2
(6.7 $\mu$m) and LW3 (15 $\mu$m) filters.  The observations were carried out in July
1996 and have been described by Serjeant et al (1997).
%\htmladdnormallink{Paper I}{../../paperi/paperi.html}).
The images have been searched
for point sources by Goldschmidt et al (1997):
%\htmladdnormallink{Paper II}{../../paperii/paperii.html}): 
a total of 15 sources were found in the
central HDF area at 6.7 $\mu$m,
 and 5 at 15 $\mu$m, of which 6 and 4, respectively, are from complete and reliable sub-samples.  
A further 27 sources were found in the 
flanking fields around the HDF.  The resulting source-counts have been
discussed by Oliver et al (1997) 
%\htmladdnormallink{Paper III}{../../paperiii/paperiii.html})
and shown to be consistent with the strongly evolving
starburst models previously used to model the 60 $\mu$m and 1.4 GHz counts (Franceschini et al 1994, 
Rowan-Robinson et al 1993, 
Pearson and Rowan-Robinson 1996).  Associations for the 17 ISO sources in the central HDF area (2 were 
detected at both 6.7 and 15 $\mu$m) were sought with HDF galaxies 
using a likelihood method (Mann et al 1997) 
%\htmladdnormallink{Paper IV}{../../paperiv/paperiv.html}) 
and 13 credible associations were found.  In this paper we take the view that these associations
tentatively confirm the reality of those sources which are not in the reliable and complete sub-samples.
There is ambiguity about some of the associations (Mann et al 1997)
and in some case the ISO flux may be due to more
than one galaxy (this is particlularly so for the 15 $\mu$m detections).  In this paper we have assumed 
that all the flux is assigned to the galaxy with the highest likelihood.  This assumption does not have
a great effect on our overall conclusions.  For the two sources where the likelihoods did not completely
resolve ambiguities (12 36 43.0 +62 11 52 and 12 36 48.4 +62 12 15), we have conservatively chosen the 
lower redshift galaxy as the association.

In this paper we discuss the spectral energy distribution of the 13 galaxies detected by ISO in
the central HDF area  and consider the implications for star 
formation rates and the overall history of star formation in the universe.
Details of the 13 galaxies are given in Table 1.

A striking feature of the fainter HDF galaxies is their blue colours, indicative of
high redshift galaxies undergoing bursts of star formation.  This is confirmed both
by systematic analyses of the colours of the HDF galaxies (Mobasher et al 1996) and 
by studies of the morphologies of the galaxies (Abraham et al 1996), which show a high 
proportion of interacting and merging systems.  In both respects the HDF galaxies look 
like a higher redshift version of the starburst galaxies found by IRAS.  This was one
of the strong motivations for seeking observing time with ISO.  The fact that HDF galaxies
have been detected by ISO is sufficient to demonstrate that this analogy with IRAS
galaxies is highly relevant.

An Einstein de Sitter model ( $\Omega_{0}$ = 1), with a Hubble constant $H_{0}$ = 50 km/s/Mpc, has been used
throughout the paper.

\section{Spectral Energy Distributions}
For the 13 galaxies reliably detected by ISO in the HDF (Goldschmidt et al 1997, Mann et al 1997) we have modelled their
spectral energy distributions (seds) from 0.3 to 15 $\mu$m.  Spectroscopic redshifts are available for 9 
of the galaxies (Cowie 1996, Cohen et al 1996, Phillips et al 1997).  For the remaining 4 we have used photometric redshifts determined by the
method of Mobasher et al (1996).  The latter analysis has been repeated using the total magnitudes
and colours given in the STScI HDF catalogue (Williams et al 1996).  The resulting redshifts agree well with 
those determined in a small, fixed aperture by Mobasher et al (1996).

The U,B,V,I data (AB magnitudes) from HST and the J, H/K data of Cowie (1996) have been 
fitted with galaxy models from the library of Bruzual and Charlot (1993).  The near infrared data
were corrected to the same aperture as the optical data.  For wavelengths beyond 2.5 $\mu$m,
the Bruzual and Charlot models give predictions based on IRAS data for their standard stars, which
appear to generate a spurious secondary peak at 12 $\mu$m (even for the youngest starburst models).  
We have therefore replaced the Bruzual and Charlot predictions with a Rayleigh-Jeans extrapolation
beyond 2.5 $\mu$m.
Excellent fits to the U to K data for our galaxies can be
obtained using models with starburst of duration $10^{9}$ yrs, viewed at a range of subsequent
times $\tau$ = 1 to 2.4 Gyr (values of $\tau$ are given in Table 1).  In the case of the galaxies
for which we have only photometric redshifts, the good fits of the
models to the data provide support for the photometric redshifts.
Almost equally good fits could be obtained with an exponentially decreasing star formation rate 
with a time-scale of 1 Gyr.
No allowance is made for reddening at this stage.  For one galaxy (12 36 43.9 +62 11 30), the 6.7 $\mu$m
emission can be accounted for almost completely by starlight.  

For the remaining 12 galaxies, there is a clear excess of infrared radiation.  We have
considered first the possibility that we are seeing the infrared 'cirrus' in the galaxies,
emission from starlight in the galaxies absorbed by interstellar grains and reemitted in
the infrared.  The cirrus models of Rowan-Robinson (1992) have been revised to incorporate
very small grains and PAHs correctly (Efstathiou, Rowan-Robinson and 
Siebenmorgen 1997, in preparation).  For 12 of the galaxies,
there was no plausible cirrus model, because the resulting far infrared luminosity
was always at least 3 times the total optical-uv luminosity of the galaxy.  A typical
value of $L_{fir}/L_{opt-uv}$ for cirrus emission is 0.2-0.3 (Rowan-Robinson et al 1987,
Rowan-Robinson 1992).  For one galaxy (12 36 48.1 +62 14 32), in which the cirrus model 
gave a far infrared luminosity 
comparable to that seen in the optical and uv, we accepted the cirrus model fit (see Fig 1).  

Figure 1 shows fits of the standard starburst model of Efstathiou et al
(1997) to the infrared spectral energy distributions of the remaining 11 ISO-detected HDF 
galaxies.  The model is a
development of the earlier starburst model of Rowan-Robinson and Efstathiou (1993), with a proper treatment of very small grains and PAHs, and gives an excellent
fit to the spectra of M82 and the starburst galaxy NGC6090 studied by ISO (Acosta-Pulido et al 1996).
Although in most cases we have only a single mid-infrared data point, it is
satisfactory that in the case where we have detections at both 6.7 and 15 $\mu$m
, the model fits both data points well.  Where only upper limits are available at one of the ISO
wavelengths, these are generally consistent with the predictions of the model (for objects
4 and 5 in Table 1, the upper limits at 15 $\mu$m may imply that the model parameters
need adjustment, that an alternative model, eg a dusty AGN torus, may be needed, or that
the 6.7 $\mu$m detections is unreliable).  Even more impressive is the fit for 3 galaxies to the VLA
detections by Fomalont et al (1996), for which we have assumed in the model the standard radio-far ir 
correlation ( S(60 $\mu$m)/S(1.4 GHz) = 90 ) and a radio spectral index of 0.8.  This
supports the idea that we really are seeing dust emission from starbursts with ISO.  
In most other cases the models are consistent with the 1.4 GHz upper limit, taken as 12.2 $\mu$Jy
(Fomalont et al 1997) ( but for objects 4, 5 and 10, the radio limits lie significantly below 
the predictions of the starburst model).  
If we look at the morphologies of the galaxies in our sample, the 2 non-starburst galaxies are
both ellipticals.  Of the 11 galaxies whose seds we fit as starbursts, 8 are spirals, 2 are interacting
pairs and 1 is a galaxy (possibly spiral)  with a prominent nucleus (12 36 46.4 +62 14 06).  Broad lines have been found 
in the optical spectrum of this galaxy and
it is possible that the 6.7 mu emission could be from a dusty torus surrounding
the AGN, rather than from a starburst.  However if this interpretation were correct, the agreement
of the radio flux with the prediction of the starburst model would be a coincidence.  There are 
several other objects in which a dusty torus model would give an equally satisfactory fit to the
observed infrared excess.  However since the covering factor by dust in AGN is generally found to be $< 0.5$
(Rowan-Robinson 1995), it is unlikely that more than one object in our sample is AGN with its dust
torus seen edge-on.  The source-count models discussed by Oliver et al (1997) predict that the fraction
of AGN expected in 6.7 and 15 $\mu$m samples should be small.  We therefore assume that the infrared
excess in the remaining 11 objects is from starbursts rather than from AGN with dusty tori.

Table 1 gives the inferred rest-frame values of $\nu L_{\nu}$ at 0.3, 0.8, 15 and 60
$\mu$m.  It can be seen that $L_{60}/L_{0.8}$ ranges from 2 to 200, and $L_{60}/L_{0.3}$ ranges
from 3 to 1000.  Thus the
implication of the ISO detections is that, at least for the detected galaxies,
the bulk of the bolometric luminosity of the galaxies is emitted at far infrared
wavelengths.  The interpretation of this is similar to that for the starburst
galaxies found in IRAS surveys: most massive star formation takes place in
dense molecular clouds and is shrouded from view by a substantial optical depth
in dust.  What is seen in the optical and uv represents stars formed near the
edges of clouds, so that the light from these stars can escape directly.

\begin{figure*}
\epsfig{file=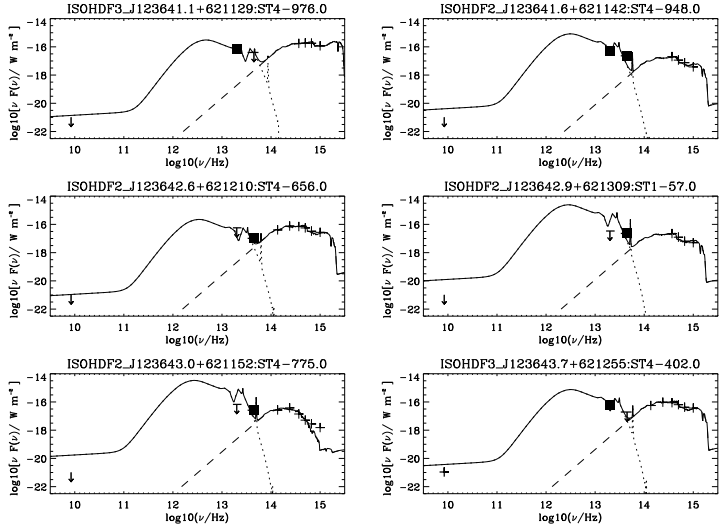,angle=0,width=10cm}
\caption{
(a) Spectral energy distributions ( $\nu F(\nu)$ in $W m^{-2}$) for galaxies of Table 1,
in order of right ascension, compared with Bruzual and Charlot (1993) models (optical and near ir) and
Efstathiou et al (1997) cirrus and starburst models (mid and far ir).  Crosses: HST, IFA and VLA data, 
filled squares: ISO data.}
\end{figure*}

\begin{figure*}
\epsfig{file=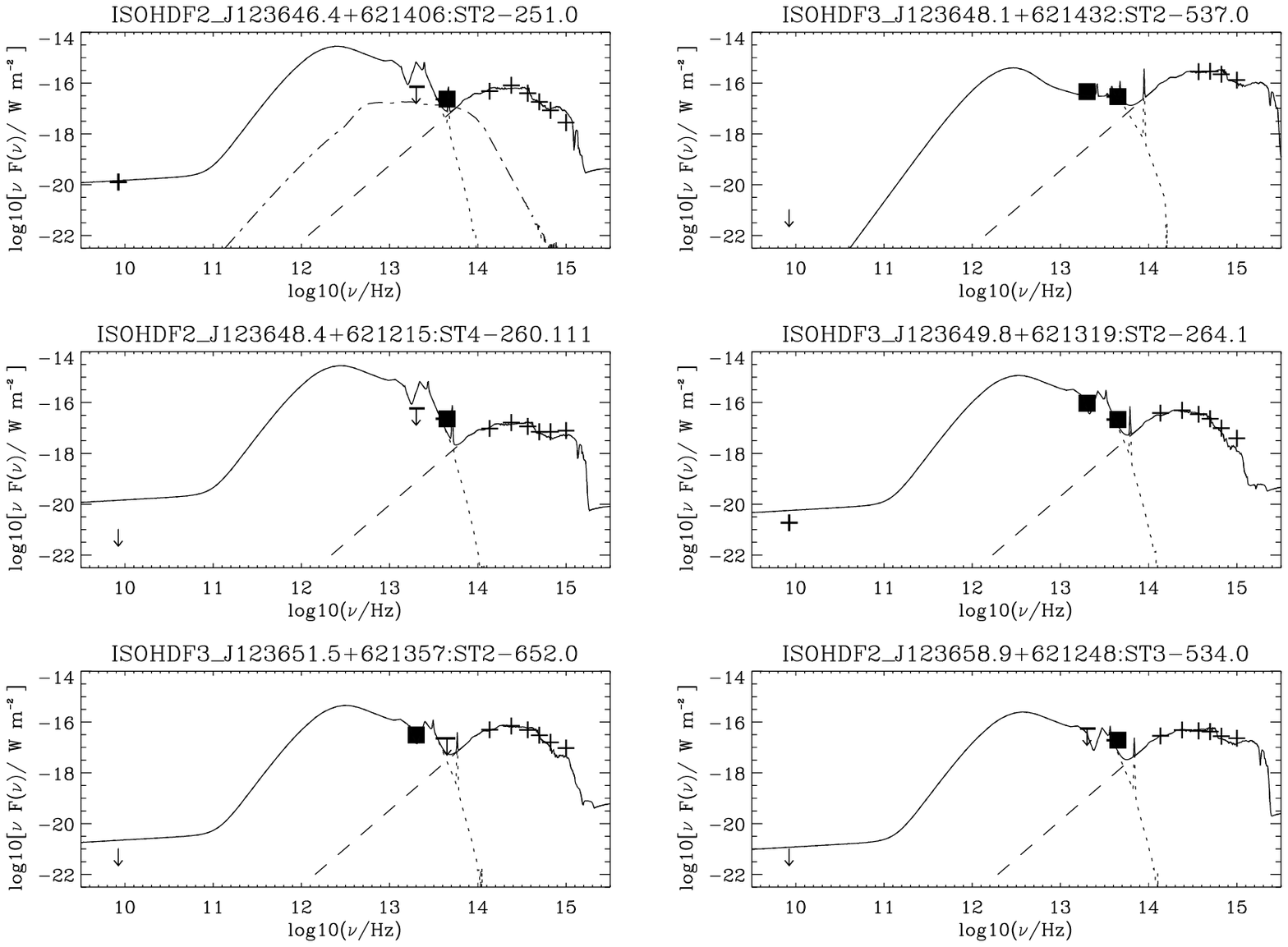,angle=0,width=10cm}
\contcaption{
(b) Caption as for Fig 1a.
}
\end{figure*}

\begin{table*}
\begin{minipage}{140mm}
\caption{60 $\mu$m luminosities and star formation rates for ISO-HDF galaxies}
\begin{tabular}{llllllllll}
 & {\bf Galaxy} & {\bf redshift} & $\tau$ & {\bf $L_{0.3}/L_{\odot}$} & {\bf $L_{0.8}/L_{\odot}$} & {\bf $L_{15}/L_{\odot}$} & {\bf $L_{60}/L_{\odot}$} & {\bf $(\dot{M}_{*,all}/M_{\odot})$} & notes \\
 & (ISOHDF)& & (Gyr) & & & & & x$\phi^{-1}$ & \\
 & & & & & & & \\
1 & 12 36 41.1  +62 11 29 & (0.047) & 1.02 & 2.8x$10^{8}$& 4.0x$10^{8}$ & 2.1x$10^{8}$ & 7.6x$10^{8}$ & 0.20 & S sb  c \\
2 & 12 36 41.6  +62 11 42 & 0.585 & 1.14 & 2.8x$10^{9}$ & 7.3x$10^{9}$ & 1.1x$10^{11}$ & 3.9x$10^{11}$ & 101 & I sb b,d \\
3 & 12 36 42.6  +62 12 10 & 0.454 & 1.14 & 6.9x$10^{9}$ & 1.8x$10^{10}$ & 1.7x$10^{10}$ & 6.0x$10^{10}$ & 16 & S sb b \\
4 & 12 36 42.9  +62 13 09 & (0.74) & 1.14 & 5.5x$10^{9}$ & 1.4x$10^{10}$ & 5.3x$10^{11}$ & 1.9x$10^{12}$ & 495 & S sb b\\
5 & 12 36 43.0  +62 11 52 & (0.82) & 2.4 & 3.4x$10^{9}$ & 3.0x$10^{10}$ & 8.9x$10^{11}$ & 3.2x$10^{12}$ & 840 & S sb a \\
6 & 12 36 43.7  +62 12 55 & 0.558 & 1.14& 1.4x$10^{10}$ & 3.6x$10^{10}$ & 8.6x$10^{10}$ & 3.1x$10^{11}$ & 80 & I sb c,e \\
7 & 12 36 43.9  +62 11 30 & 1.01 & 3.5 & 1.75x$10^{10}$ & 1.35x$10^{11}$ & 2.0x$10^{10}$  & 1.4x$10^{11}$ & - & E sl b,e\\
8 & 12 36 46.4  +62 14 06 & 0.960 & 1.28 & 2.2x$10^{10}$ & 8.1x$10^{10}$ & 1.1x$10^{12}$ & 3.9x$10^{12}$ & 1010 & E sb a,e \\
9 & 12 36 48.1  +62 14 32 & (0.023) & 1.14 & 6.0x$10^{7}$ & 1.5x$10^{8}$ & 1.8x$10^{7}$ & 1.26x$10^{8}$ & - & E cirr a,c \\
10 & 12 36 48.4  +62 12 15 & (0.778) & 1.02 & 4.5x$10^{9}$ & 1.15x$10^{10}$ & 6.8x$10^{11}$ & 2.45x$10^{12}$ & 640 & S sb a \\
11 & 12 36 49.7  +62 13 15 & 0.475 & 1.28 & 1.75x$10^{9}$ & 1.26x$10^{10}$ & 9.5x$10^{10}$ & 3.4x$10^{11}$ & 88 & S sb a,c,e \\
12 & 12 36 51.5  +62 13 57 & 0.557 & 1.14 & 4.7x$10^{9}$ & 2.7x$10^{10}$ & 5.2x$10^{10}$ & 1.86x$10^{11}$ & 48 & S sb d \\
13 & 12 36 58.9  +62 12 48 & 0.320 & 1.02 & 2.4x$10^{9}$ & 6.0x$10^{9}$ & 9.0x$10^{9}$ & 3.2x$10^{10}$ & 8 & S sb b\\
\end{tabular}
\medskip

S -   spiral

E -   elliptical

I -   interacting pair

sb -   sed fitted with starburst model

cirr -   sed fitted with cirrus model

sl -   sed fitted with starlight

a -   detected at 6.7 $\mu$m, from reliable and complete sub-sample

b -   detected at 6.7 $\mu$m, from supplementary list

c -   detected at 15 $\mu$m, from reliable and complete sub-sample

d -   detected at 15 $\mu$m, from supplementary list

e -   detected at 1.4 GHz (Fomalont et al 1997)
\end{minipage}
\end{table*}

\section{Star formation rate}
A number of authors have discussed how the star formation rate in a galaxy can be inferred 
from its optical, ultraviolet or far infrared luminosity.  
Scoville and Young (1983)
estimated the star formation rate of O,B,A stars ($M > 1.6 M_{o}$) from the total far 
infrared luminosity of galaxies, 
implicitly assuming a burst of star formation lasts $10^{9}$ yrs, finding

$\dot{M}_{*,OBA}$ = 7.7x$10^{-11} L_{fir}/L_{o}$ .				(1)

Thronson and Telesco (1986) used
a Salpeter IMF to give rates of star formation of all stars and of
OBA stars, averaged over the past 2x$10^{6}$ yrs, per unit far infrared luminosity:

$\dot{M}_{*,OBA}$ = 2.1x$10^{-10} L_{fir}/L_{o}$ .				(2)

$\dot{M}_{*,all}$ = 6.5x$10^{-10} L_{fir}/L_{o}$ .				(3)

They attribute the fact that (2) is a factor of 3 higher than (1) to the different assumptions
about the duration of the burst.
They also give the scaling factors for the star formation rate 
between the different IMFs and lower mass limits:

M/L (Miller-Scalo, 100, 0.1) : M/L (M-S, 100, 1.6) : M/L (Salpeter, 100, 0.1): M/L (Salpeter, 100, 1.6)

= 10.2 : 4.0 : 3.1 : 1

More recently Madau et al (1996) have calculated total star formation rates and heavy element
production, $\dot{M}_{Z}$, from the ultraviolet luminosity densities at 
2800 $\AA$, using
a Salpeter IMF and
the evolutionary models of Bruzual and Charlot (1993).  The figures they give are equivalent to

$\dot{M}_{*,all}$ = 5.3x$10^{-10} L_{2800}/L_{o}$ .				(4)

$\dot{M}_{*,all}$ = 42 $\dot{M}_{Z}$						(5)

To convert from 60 $\mu$m luminosity to star formation rate, we assume 
that a fraction $\epsilon$ 
($\simeq$ 1) of the optical and uv energy emitted in a starburst is absorbed by dust
and emitted in the far infrared, so that 

$L_{bol,fir}$ = $\epsilon L_{bol,opt-uv}$.					(6)

The bolometric correction at 2800 $\AA$ for the 1 Gyr starburst models of Bruzual and Charlot (1993), when
viewed at early ages, is 3.5 
and those at 15 and 60 $\mu$m for the dusty starburst model of Rowan-Robinson and Efstathiou (1993) are 6.0 and 1.7,
respectively, so using (4) and (6):

$\dot{M}_{*,all} /[L_{60}/L_{o}]$ = 2.6 $\phi/\epsilon$ x$10^{-10}$			(7)

$\dot{M}_{*,all} /[L_{15}/L_{o}]$ = 9.3 $\phi/\epsilon$ x$10^{-10}$			

where the factor $\phi$ incorporates (1) the correction from a
Salpeter IMF to the true IMF (x3.3 if the Miller-Scalo IMF is the correct one), (2) a correction if
the starburst event is forming only massive stars (x 1/3.1 if only O,B,A 
stars, $> 1.6 M_{o}$, are
being formed).  This estimate, which is now based on detailed starburst models for the optical-uv radiation
and proper radiative transfer models for the far infrared emission, is a factor 1.9 higher than that
of Scoville and Young (1983, eqn (1) above) and a factor 0.7 times that of Thronson and Telesco (1986, eqn
(2) and (3) above).

\section{Star formation rates for ISO-HDF galaxies}
Table 1 gives the inferred 15 and 60 $\mu$m luminosities, and star formation rates based on eqn (7), for
the 11 HDF starburst galaxies detected by ISO.  The star formation rates range (with one exception) 
from 8-1000 $\phi M_{o}$ per yr.  The galaxies detected by ISO are forming stars at a 
prodigous rate compared with nearby normal spirals.  Although star formation rates based only on
6.7 and 15 $\mu$m detections are bound to be rather uncertain, because most of the energy is emitted
at much longer wavelengths, it is clear that the star formation rates deduced
from the uv fluxes detected by HST are a severe underestimate for these galaxies.  

It is of course of interest to ask whether these star formation rates can be typical of all the
galaxies in the HDF.  For 2 of the galaxies detected by ISO, the 6.7 and/or 15 $\mu$m flux is
consistent with emission from starlight and/or cirrus and there is no evidence for a luminous starburst.  For
other bright galaxies in the HDF, the non-detection by ISO gives a significant upper limit on
any excess far infrared emission.  For these galaxies the estimates of star formation rate from
the uv flux will be correct.
However we can not rule out the possibility that for a significant fraction of the fainter HDF galaxies,
particularly those with $z > 1$, the presence of a strong far infrared excess is the norm rather
than the exception.  When we see star formation within our Galaxy or in other nearby galaxies,
the bulk of the massive stars (which produce all the heavy elements and most of the ultraviolet
light) are formed within dense molecular clouds behind a high optical depth in dust.  The starburst
galaxies detected by IRAS emit most of their radiation at far infrared wavelengths.  It is a reasonable
expectation that as we look back to epochs when the bulk of the stars in a galaxy are formed,
that this too will take place within dense clouds of molecules and dust and be primarily a
far infrared phenomenon.  Of course eventually we will see back to epochs when the very first
stars form, when little or no heavy elements or dust are present, and 
star formation will be entirely an optical and uv phenomenon.
However this transparent phase may last no more than a few percent of the main star
formation phase, say $10^{7} - 10^{8}$ yrs, and be confined to very high redshifts ( $> 3-5$).

\section{Infrared luminosity-density and the history of star formation}
Madau et al (1996) have used the Canada-France Redshift Survey (Lilly et al 1996) and HDF data to calculate the history of
star formation and heavy element generation, under the assumption
that the uv gives a complete view of the star formation that is occurring.  Integrating 
over the star formation density 
as a function of redshift, they conclude that all the heavy elements associated with the
visible matter in galaxies can be generated.  However if their calculated baryonic density is
converted to a value for $\Omega$, a value of 0.0035 is obtained, only 7 $\%$ of the baryonic 
density of 0.05, for an assumed $H_{o}$ = 50, derived from cosmological nucleosynthesis of the light 
elements (Walker et al 1991).  It is not unreasonable to assume that some of the remaining
93 $\%$ or baryons has participated in star formation and heavy element production.  For example, the
hot gas in clusters is known to have a heavy element abundance of at least 1/3rd of solar.

We first estimate the far infrared luminosity density for the luminosity function derived
from IRAS 60 $\mu$m data.
Oliver et al (1997) have shown that the 6.7 and 15 $\mu$m source-counts
are consistent with the strongly (luminosity-)evolving starburst models which we have used
to fit (1) the redshift distribution in IRAS redshift surveys (Saunders et al 1990, Oliver et
al 1995), (2) the 60 $\mu$m source-counts (Pearson and Rowan-Robinson 1996), (3) the far
infrared background, including the claimed detection using FIRAS data from COBE
 by Puget et al (1996)
(Pearson and Rown-Robinson 1996, Rowan-Robinson and Pearson 1996), (4) the sub-mJy 1.4 GHz 
radio counts (Rowan-Robinson et al 1993, Hopkins et al 1996).

The solid curve in Fig 2 shows the luminosity-density at 60 $\mu$m as a function of redshift.  
For $z < 0.3$ this
is directly derived from IRAS galaxy redshift surveys (the luminosity function given in line
(23) of Table 3 of Saunders et al 1990).  The extrapolation to higher redshift is the luminosity
evolution model used in Pearson and Rowan-Robinson (1996), Rowan-Robinson et al (1993), and
Oliver et al (1997) to fit the deep 60 $\mu$m and 1.4 GHz source-counts, for which

$L_{*}(z)$ = $(1+z)^{3.1}, z < 2$,						(8)

$L_{*}(z)$ = $3^{3.1}, 2 < z < 5$.

Using eqn (7) we can convert this to a density of star formation, and integrate to derive a
total mass-density in stars or in heavy elements.  We find

$\Omega_{*}$ = 0.008 $h^{-2}_{50}$ $\phi$,		$\Omega_{Z}$ =0.00019 $h^{-2}_{50}$.		(9)

These values are not unreasonable.   They require that twice as much star formation as has been inferred
by Madau et al (1996) from the uv integrated light has taken place shrouded by dust.  The total fraction of
baryonic matter that has participated in star formation would be of order 20 $\%$, with about
1/3rd of the 
resulting heavy elements now residing in the luminous parts of galaxies.  The remainder could be in baryonic
objects in the halos of galaxies or in intergalactic gas (including the hot X-ray emitting gas in clusters).
In fact evolutionary rates appreciably steeper than that assumed in eqn (8) can probably not be ruled out
at this stage.  If the star-forming galaxies we have detected with ISO are typical of the fainter HDF galaxies,
then we may require that more than 50 $\%$ of baryons have participated in star-formation and heavy element production, 
presumably with a truncated IMF so that most of the baryons now reside in dark remnants.  
Similar conclusions are reached if we use the model for evolution of infrared galaxies of
Franceschini et al (1994, 1997 in preparation), shown as a broken line in Fig 2.  

We have also estimated the contribution to the 60 $\mu$m luminosity-density implied directly by the
ISO-HDF starburst galaxies.  There are 5 starburst galaxies in the redshift bin 0.4-0.7 and 3 in the redshift bin 0.7-1.0
(omitted the galaxy with broad lines).  Estimating the volume of the universe sampled by the HDF survey
we find contributions of 6.0 $\pm$2.0x$10^{8}$ and 2.6 $\pm$1.5x$10^{9} L_{o} Mpc^{-3}$ for the
redshift ranges 0.4-0.7 and 0.7-1.0 respectively.  These estimates take no account of sources fainter
than the ISO limit and they are subject to any uncertainty in the ISO calibration (probably a factor of 50%
either way), as well as the considerable uncertainty associated with extrapolating from 6.7 and 15 $\mu$m
to 60 $\mu$m, so must be seen as very preliminary.  They appear to be significantly
higher than the predictions for evolution of the form (8) (the solid line in Fig 3) by factors of 5 and 10
respectively.  This may imply that the evolution of starburst galaxies is steeper than assumed in eqn (8)
for 0 $<$ z $<$ 1.  Alternatively our models may overestimate the 60 $\mu$m luminosities for at least some of
the galaxies, for example because the 6.7 and 15 $\mu$m radiation comes from dust tori around AGN.

We have also shown in Fig 2 one of the more extreme of the models of Pei and Fall (1995, dotted line).  This
illustrates that the luminosity density estimated from the ISO-HDF galaxies is not at odds with current data 
on the number-density of quasar absorption-line clouds or the observed heavy element abundance at high redshift.
However for this model $\Omega_*=0.032 h^{-1}_{50}\phi$, $\Omega_{\rm
Z}=0.00076 h^{-1}_{50}$, which would probably imply that only higher
mass stars were being formed in the ISO-HDF galaxies.

Support for the idea that stronger evolution than (8) is required for the starburst galaxy population comes
from the fit to the ISO counts by Oliver et al (1997).  The Pearson and Rowan-Robinson (1996) model fails to
predict a strong enough contribution to the counts by starburst galaxies.  The Franceschini et al (1994) model,
involving a strongly evolving starburst population in elliptical galaxies, appears to give a better fit to
the ISO counts.  This idea can be tested by the deep 90 $\mu$m surveys which we and others are carrying out
with ISO.
It will also be interesting to observe the HDF galaxies detected by ISO at submillimetre wavelengths, for example
with SCUBA on the JCMT.

\begin{figure*}
\epsfig{file=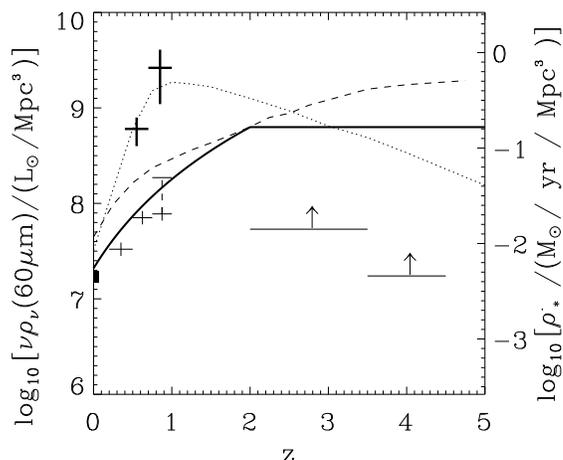,angle=0,width=8cm}
\vspace*{0.4cm}
\caption{
The luminosity density at 60 $\mu$m as a function of redshift (left-hand scale) and the star-formation
rate (right-hand scale, from eqn (7) assuming $\phi$ = 1).  
The solid curve is derived from the IRAS 60 $\mu$m luminosity function of Saunders et al (1990, assuming evolution
of the form (8).  The broken curve is the model of Franceschini et al (1994, 1997)
and the dotted curve is the infall model of Pei and Fall (1995) with 
$\Omega_{g}$ = 8x$10^{-3} h^{-1}$.
The thick crosses are the estimates of luminosity density derived directly from the ISO data for the galaxies 
of Table 1.
Also shown (thin crosses and lower limits) are the star formation rates
derived from the ultraviolet (2800 $\AA$) luminosity density by Madau et al (1996).   
}
\end{figure*}

\section{Conclusions}
We have modelled the spectral energy distributions of the 13 galaxies reliably associated with ISO 
sources detected at 6.7 and$/$or 15 $\mu$m. 
For 2 galaxies the emission detected by ISO is
consistent with being starlight or normal infrared 'cirrus' in the galaxies.  For the remaining 11 galaxies there
is a strong mid-infrared excess, which we interpret as emission from dust associated with a strong
starburst.  In 3 cases the starburst model appears to be
confirmed by the good agreement of the the predicted radio flux with that detected by Fomalont et al
(1996).
Inferred rest-frame luminosities ($\nu L_{\nu}$) at 0.3, 0.8, 15 and 60 $\mu$m are given and $L_{60}/L_{0.3}$
ranges from 3 to 1000 for the 11 galaxies.  Thus most of the the bolometric luminosity of the galaxies is
predicted to emerge at far infrared wavelengths.

We give a new discussion of how the star formation rate can be deduced from the far infrared luminosity
and derive star formation rates of 8-1000 $\phi M_{o}$ per yr, where $\phi$ takes account of the
uncertainty in the initial mass function (=1 for Salpeter IMF).  The HDF galaxies detected by ISO
are clearly forming stars at a prodigous rate compared with nearby normal galaxies. We discuss the
implications of our detections, and of the IRAS 60 $\mu$m luminosity function and evolution, for the
history of star and heavy element formation in the universe.  We conclude that at least 20 $\%$ of
baryons must have participated in star formation.

\section*{
Acknowledgements} 
This paper is based on observations with ISO, an ESA project, with instruments
funded by ESA Member States (especially the PI countries: France,
Germany, the Netherlands and the United Kingdom) and with the
participation of ISAS and NASA.
We thank the referee, Harry Fergurson, for comments and suggestions
which enabled us to improve this paper.
This work was supported by PPARC
(Grant no. GR/K97828) and by the EC TMR Network Programme (Contract no. FMRX-CT96-0068).

\end{document}